\documentclass[12pt]{article}

\usepackage{paper2e}
\usepackage{mydefs2e}

\newcommand{\LIR}{\Lambda_{\rm IR}}
\newcommand{\Pt}{{\tilde{\Phi}}}
\newcommand{\Ft}{{\tilde{F}}}
\renewcommand{\d}{\partial}
\newcommand{\dd}{\raisebox{1.2pt}{$\stackrel{\raisebox{-1pt}%
{$\scriptscriptstyle\leftrightarrow$}}{\d}$}}

\begin{document}

\begin{titlepage}
\preprint{BA-05-103\\KEK-TH-1044}

\title{Gauge Mediation from Emergent Supersymmetry}
\author{Hock-Seng Goh%
\footnote{\tt hsgoh@physics.arizona.edu}}%
\address{Department of Physics, University of Arizona, Tucson, AZ 85721, USA.}

\author{Siew-Phang Ng%
\footnote{\tt spng@bartol.udel.edu}}%
\address{Bartol Research Institute, University of Delaware,
Newark, Delaware 19716, USA.}

\author{Nobuchika Okada%
\footnote{\tt okadan@post.kek.jp}}
\address{Theory Division, KEK, Oho 1-1, Tsukuba, Ibaraki 305-0801,
Japan.\\
Department of Particle and Nuclear Physics, The Graduate
University for Advanced Studies (Sokendai), Oho 1-1, Tsukuba,
Ibaraki 305-0801, Japan.}

\begin{abstract}
We explore the possibility of gauge mediation in a paradigm
whereby supersymmetry is posited to be an accidental symmetry of
Nature and the Standard Model fields are composite bound states
that emerge from a conformal field theory. The resultant effective
theory can, through sequestering and conformal dynamics, exhibit
most of the properties of low energy supersymmetry breaking while
averting a number of cosmological and astrophysical constraints of
the traditional framework of gauge mediation via dynamical
supersymmetry breaking. Of particular phenomenological interest is
that in our scenario, the gravitino is superheavy, the neutralino
LSP is a viable candidate for cold dark matter and the flavor
changing neutral currents are constrained to be, at the very
minimum, only an order of magnitude below current experimental
bounds.
\end{abstract}

\end{titlepage}

\section{Introduction}

One of the most pressing and immediate concerns confronting
theoretical high energy physics today is the question of what lies
beyond the Standard Model (SM). Arguably, the most attractive
scenario of physics beyond the SM is the existence of
supersymmetry (SUSY). As is clear from the absence of
supersymmetry in our world, it is manifestly broken. The breaking
of supersymmetry in our sector, which shows itself in the
sparticle spectrum, is dictated by the mediation mechanism or
mechanisms as the case may be.

Gauge mediation\cite{GMSB}, by which our sector feels SUSY
breaking through messenger fields charged under the SM gauge
groups, ranks among the leading candidates for transmission of
supersymmetry breaking. A simple and predictive mechanism, it
gives flavor-independent contributions to soft masses that is
proportional to the order parameter $\frac{F}{M}$ where $F$ and
$M$ are respectively the F-term and the scalar vacuum expectation
value (vev) of some non-SM field that couples to the messenger
fields. The question then arises as to how one might obtain $F$
and $M$. In the conventional picture of gauge mediation, $F$ and
$M$ are obtained when SUSY is spontaneously broken by
non-perturbative dynamical effects in a theory that possesses
tree-level supersymmetric vacua, i.e. by dimensional
transmutation. Dynamical supersymmetry breaking
(DSB)\cite{dynSUSY}, as this mechanism is known, ensures the
stability of the gauge hierarchy while at the same time evading
the supertrace theorem that rules out communication of SUSY
breaking through tree-level renormalizable couplings.
Unfortunately, the phenomenology arising from this scenario or any
other within the conventional framework usually leads to a
gravitino lightest supersymmetric particle (LSP) as well as other
light moduli fields\footnote{In string or higher-dimensional
supergravity theories, there generically exists a large number of
flat directions which typically take on masses of the order of the
gravitino mass, $m_{3/2}$, once supersymmetry is broken.} that are
plagued by a plethora of cosmological problems
\cite{MoroiYanagida}.

Perhaps we need to take a step back and look at the theoretical
assumptions of the above scenario, and see if there is a more
compelling and natural framework in which to consider gauge
mediation. One intriguing possibility is the ``Supersymmetry
without Supersymmetry'' paradigm\cite{SwoS} where SUSY is posited
to be an accidental symmetry of Nature(see also
\Refs{GherghettaPom, Kaplan}). This scenario, which constitutes the theoretical
underpinnings of the present paper, starts off with completely
non-supersymmetric theories that flow, in the Wilsonian sense, to
more supersymmetric ones. Though possessing the interesting
property that the gravitino and potential cosmological moduli
fields are superheavy in this scenario, ultimately the theory does
not solve the supersymmetric flavor problem. This seems to be a
generic problem with previously considered theories of this class,
i.e. the ``Almost No-scale Supergravity'' scenario\cite{Noscale}.
In this paper, we shall present an explicit model based on gauge
mediation within the ``Supersymmetry without Supersymmetry''
paradigm that addresses the issues and shortcomings we have
considered above. Before we give a summary of the results, a short
discussion of the general paradigm is warranted. We shall, in the
course of our discussion, be utilizing the AdS-CFT
correspondence\cite{AdSCFT} to extract useful insights from both
perspectives.

Let us begin our discussion by questioning the implicit assumption
that lies at the heart of the conventional SUSY picture. What if
SUSY is not a fundamental symmetry of Nature? Even if it were, let
us break it at the Planck scale. The fermion-boson splitting, and
by extension the natural mass of the gravitino and the moduli
fields, would then be Planckian in magnitude. Ostensibly, it would
seem that we are re-introducing the hierarchy problem and we are
no better off than when we started. However, this could be evaded
if the Higgs fields as well as the rest of the SM are composite
particles that ``emerge'' from the conformal sector. The $M$ in
the order parameter is then related to the compositeness scale of
the SM while $F$ would be the amount of SUSY breaking experienced
by the SM fields. As we are considering Planck-sized fundamental
SUSY breaking, $\sqrt{F}$ could in principle be much larger than
$M$ and aside from not giving the correct observed low-energy SUSY
breaking, would also break the SM gauge symmetries. However, the
conformal sector fields can naturally conspire, by the
sequestering mechanism \cite{ConfSeq}, to systematically
ameliorate the effects of high-energy supersymmetry breaking to
give a theory that for most intents and purposes, exhibits the
appearance of low-energy SUSY breaking, i.e. Higgs, gaugino and
squark masses of $\sim O(100\mathrm{GeV})$.


What kind of conformal sector do we need to achieve
sequestering\footnote{Note that the sense in which we are
employing the word sequestering is that the SUSY breaking effects
are highly suppressed (which is crucial as we have Planckian
breaking of SUSY) in our sector. It does not necessarily mean that
anomaly mediation is the dominant contributor to SUSY breaking in
the sparticle spectrum. Indeed as we shall see, anomaly mediation
remains a sub-dominant effect in this scenario.}? First and
foremost, it has to be strongly coupled such that the operators of
the conformal field theory (CFT) have order one corrections to the
anomalous dimensions which render all supersymmetry breaking
operators sufficiently irrelevant such that their effects will
become suppressed at low energies. The other ingredient that must
be added is that there is a spontaneous breaking of the conformal
symmetry at some intermediate scale so that most of the composite
bound states (``mesons'' and ``baryons'') acquire intermediate
masses and hence can be integrated out. What remains are the
vestigial light composite (or emergent, if you like) degrees of
freedom that form an effective low energy theory that has a
gauge mediation messenger sector plus a nearly supersymmetric
extension of the Standard Model, with either MSSM or the NMSSM
being possibilities. This would then lead to a theory that is free
of the cosmological problems of traditional gauge mediation while
preserving the solution to the hierarchy problem.


Obviously, a glaring omission of the above discussion is how we
are to explicitly realize this gauge mediation from emergent
supersymmetry scenario. It is a highly non-trivial process to
construct a strongly-coupled CFT possessing all the above
properties. We could, however, exploit the AdS-CFT correspondence
\cite{AdSCFT} to construct a weakly-coupled fully-calculable
five-dimensional AdS dual of the strongly-coupled four-dimensional
CFT that would have exactly the same physics as described above.
The basic picture is that we have a Randall-Sundrum (RS) type
setup \cite{RS1} where the UV brane (identified as the Planck
scale on the CFT side) has $\sim O (1)$ SUSY breaking while the
bulk and IR brane (the compositeness scale of the CFT) are
supersymmetric, insofar as the classical action is concern. The
fact that we have a non-supersymmetric UV brane leads to SUSY
breaking throughout the entire volume but at tree-level, the above
conditions have to be chosen so as to match the superconformal
limit of the CFT which would correspond to a restoration of SUSY
on the IR brane as it is taken to infinity. This physical
separation of the two sectors can suppress the SUSY breaking seen
by the visible sector but it is an insufficient, albeit necessary,
condition for sequestering. To accomplish sequestering on the AdS
side, we require the absence of light bulk scalars which, on the
CFT side, correspond to an absence of relevant
operators(see
Ref.\cite{Strassler} for possible string-theoretic realizations). The SUSY breaking thus transmitted to the IR brane by
the massive bulk scalars would be exponentially suppressed. The
conformal symmetry breaking can then be realized on the AdS
through bulk scalar dynamics that stabilizes the radius of the
extra dimension. In the language of the CFT, the irrelevant
operators dynamically break the conformal symmetry. For the
present paper, we will be employing the racetrack mechanism in
Ref. \cite{SwoS} whereby the effective potential arising from two
bulk scalars of nearly the same mass can terminate the conformal
dynamics. The final step is the usual inclusion of a SM-charged
messenger sector to couple to the bulk scalars and, as we shall
show, through it impart the largest contribution of SUSY breaking
to the visible sector.

The end result is that we have a phenomenologically viable theory
that has among other things, a superheavy gravitino and moduli
fields that decouple entirely from the low energy physics and thus
avoid cosmological and astrophysical constraints. In our model,
the lightest neutralino is the LSP and assuming R-parity is
conserved, is a candidate for cold dark matter. Because of the
nature of the bulk (in CFT language, CFT states) SUSY breaking
mediation mechanism, the FCNC contributions are not highly
suppressed. The ratio of flavor non-diagonal to flavor diagonal
contributions is at the very minimum only $\sim 10^{-3}$ and could
be ruled out in future experiments. This is a completely different
prediction from normal gauge mediation models where the FCNCs are
negligible due to the extremely tiny ratio between the scale of
SUSY breaking to $M_P$.


The rest of the paper is organized as follows. In Section 2, we
present the setup and the equations of motion. Details of the
general solutions are available in the Appendix. In Section 3, we
calculate the various contributions to SUSY breaking in the low
energy theory thus enabling us to eliminate the models from
Section 2 that are non-viable. In Section 4, we construct an
explicit model realizing gauge mediation from emergent SUSY. We
also consider radius stabilization and the 4-D effective low
energy theory. This culminates in a discussion of the sparticle
spectrum and phenomenology arising from this class of theories as
well as its differences with conventional gauge mediation. Section
5 contains our conclusions.

\section{Setup and General Solutions}

We follow the framework of Ref.\cite{SwoS} by having a
Randall-Sundrum (RS) model \cite{RS1} with a 5-D spacetime is
compactified on a $S^1 / Z_2\times Z_2$ orbifold, with metric
\beq[RSmetric] ds^2 = e^{-2 \si(y)} \eta_{\mu\nu} dx^\mu dx^\nu +
d y^2, \eeq where $y$ is a periodic variable with period $4\ell$,
and $\si(y)$ is (+,+) under the $Z_2\times Z_2$.

With the addition of a hypermultiplet, the action is then given by
\cite{LS2},
 \beq S = -\frac{M_5^3}{k}\myint d^4 x  \myint d^4\theta\,
(\om^\dagger \om - \varphi^\dagger \varphi)+\myint d^4
x\int_0^\ell d y\, \scr{L}_{5d-hyp}, \eeq where the radion chiral
multiplet and the conformal compensator respectively contain
\beq \om &= e^{-k \ell} + \cdots + \theta^2 F_\om,\\
 \varphi &= 1 + \theta^2
F_\varphi. \eeq and the hypermultiplet Lagrangian\footnote{We have
written the action in terms of the two-sided derivative $\Pt \dd_y
\Phi = \Pt \d_y \Phi - (\d_y \Pt) \Phi$. Also, our y-integration
is defined as
$\int_0^\ell=\int_{\epsilon}^{\ell-\epsilon}+\frac{1}{2}(\int_{-\epsilon}^{\epsilon}+\int_{\ell-\epsilon}^{\ell+\epsilon})$.}
is given by \cite{MartiPom}, \beq[hyperL] \bal \scr{L}_{5d-hyp} &=
\myint d^4\theta\,e^{-2\si} (\Phi^\dagger \Phi + \Pt^\dagger \Pt)
+ \left[ \myint d^2\theta\, e^{-3\si} \left( \sfrac 12 \Pt \dd_y
\Phi + c \si' \Pt \Phi \right) + \hc \right]
\\
& \qquad - \de(y) U(\Phi,\Pt, F,\widetilde{F}) + \de(y - \ell)\,
\om^3 \left[ \myint d^4\theta\, W(\Phi,\Pt) + \hc \right].
\eal\eeq

It is useful at this point to recall the AdS-CFT correspondence
\cite{AdSCFT} which relates the mass of the bulk scalars, $m$,
with the dimension of the CFT operator, $d$, through the following
equation, $d = 2 + \sqrt{4 + m^2 / k^2}$. The bulk masses of the
scalars of the hypermultiplet above are given by\beq
m_{\Phi,\Pt}^2 = k^2 (c \mp \sfrac 32)(c \pm \sfrac 52) \eeq
Hence, the dimensions of the operators associated with the scalar
components of $\Phi$ and $\Pt$ are \beq \dim(\scr{O}_{\Phi, \Pt})
= 2 + | c \pm \sfrac 12 |. \eeq As we require the operators
associated with both $\Phi$ and $\Pt$ be irrelevant to achieve
sequestering, we are therefore constrained to $|c| > \sfrac 52$.
This, as we shall see, does not mean we have to analyze a greater
number of unique models as our orbifold parity conditions are
capable of absorbing either sign of $c$.

We can now obtain the equations of motion which we have explicitly
written out in component form for clarity\footnote{One may also
write them in terms of superfields for a more compact form.}.
 \beq \bal
    e^{-3\si}\left[\d_y + (c - \sfrac 32) \si' \right] F &= \de(y) \frac{\d U}{\d \widetilde{\Phi}} - \de(y - \ell) \om^3 \frac{\d^2
W}{\d \widetilde{\Phi}^2} \widetilde{F}\\
    e^{-3\si} \left[\d_y  - (c + \sfrac 32) \si' \right] \Ft
&= -\de(y) \frac{\d U}{\d \Phi} + \de(y - \ell) \om^3 \frac{\d^2
W}{\d \Phi^2} F\\
    e^{-3\si} \left[ \d_y  + (c - \sfrac 32) \si' \right]\Phi +
e^{-2\si} \Ft^\dagger &= \de(y) \frac{\d U}{\d \widetilde{F}} -
\de(y - \ell)\om^3 \frac{\d W}{\d \widetilde{\Phi}}\\
    e^{-3\si} \left[ \d_y  - (c + \sfrac 32) \si' \right]\Pt -
e^{-2\si} F^\dagger &= -\de(y) \frac{\d U}{\d F} + \de(y - \ell)
\om^3 \frac{\d W}{\d \Phi} \eal \eeq

The general solution (for $0<y<\ell$) is then given by \beq \bal
    F &= F_0 e^{-(c - \frac 32) \si}\\
    \Ft &= \Ft_0 \frac{\si'}{k}  e^{(c + \frac 32) \si}\\
    \Phi &= \Phi_0 e^{-(c - \frac 32) \si}
- \frac{\Ft_0^\dagger}{(2c + 1) k} e^{(c + \frac 52) \si}\\
    \Pt &=
\Pt_0 e^{(c + \frac 32) \si} - \frac{F_0^\dagger}{(2 c - 1) k}
e^{-(c - \frac 52) \si}  \eal \eeq

The jump conditions at the UV and IR branes dictate\footnote{In
the interest of generality, we have not yet specified a specific
set boundary conditions.} the following relations between bulk and
brane quantities. \beq[jumpF] \bal
    F_{\rm UV}^+- F_{\rm UV}^-&=  \frac{\d U}{\d \Pt_{\rm UV}}\\
    F_{\rm IR}^+- F_{\rm IR}^- &=  -\frac{\d^2 W}{\d \Pt_{\rm IR}^2} \Ft_{\rm IR}\\
    \Ft_{\rm UV}^+- \Ft_{\rm UV}^-&= - \frac{\d U}{\d \Phi_{\rm UV}}\\
    \Ft_{\rm IR}^+-\Ft_{\rm IR}^- &=  \frac{\d^2 W}{\d \Phi_{\rm IR}^2} F_{\rm IR}
 \eal\eeq
and
 \beq[jumpPhi] \bal
    \Phi_{\rm UV}^+-\Phi_{\rm UV}^- &=  \frac{\d U}{\d \Ft_{\rm UV}}\\
    \Phi_{\rm IR}^+-\Phi_{\rm IR}^- &= -\frac{\d W}{\d \Pt_{\rm IR}}\\
    \Pt_{\rm UV}^+-\Pt_{\rm UV}^- &= -\frac{\d U}{\d F_{\rm UV}}\\
    \Pt_{\rm IR}^+-\Pt_{\rm IR}^- &=  \frac{\d W}{\d \Phi_{\rm IR}}
\eal \eeq where we have defined $\mathrm{lim_{\epsilon\rightarrow
0}}f(\pm\epsilon)\equiv f_{\rm UV}^{\pm}$,
$\mathrm{lim_{\epsilon\rightarrow 0}} f(\ell \pm\epsilon) \equiv
f_{\rm IR}^{\pm}$, $f(0) \equiv f_{\rm UV}$ and $f(\ell) \equiv
f_{\rm IR}$.

The above results can then be used to write down the effective
potential purely in terms of brane-localized quantities.


 \beq \bal
    \textmd{V}_{\rm eff}
    =&\, \om^3 \left( -\Phi_{\rm IR}^- \Ft_{\rm IR}^- + \Pt_{\rm IR}^-
F_{\rm IR}^- + \hc \right) +U\\
    & + \left( \Phi_{\rm UV} \Ft_{\rm UV}^+-\Pt_{\rm UV} F_{\rm
UV}^+ -\Ft_{\rm UV}\Phi_{\rm UV}^++ F_{\rm UV}\Pt_{\rm UV}^++ \hc
\right)\eal \eeq Consistent orbifolding requires that the orbifold
parity of $\Pt$ be opposite to that of $\Phi$ under both $Z_2$ and
we have used this to simplify the general potential to the above
form. We can further simplify the form of the potential by
assigning to $\Ph$ either $(+,+)$, $(-,+)$, $(+,-)$ or $(-,-)$
under the $Z_{2, \mathrm{UV}} \times Z_{2, \mathrm{IR}}$. Note
that we can perform the orbifold parity reversal for all the above
jump conditions, $+\leftrightarrow -$, and also the following
transformations $c\rightarrow -c$, $\Phi \rightarrow -\Pt$ and
$\Pt \rightarrow \Phi$ to maintain the same equations of motion
and potential. This essentially reduces the number of cases to the
following four : (i) $(+,+)$ and $c>0$, (ii) $(+,+)$ and $c<0$,
(iii) $(-,+)$ and $c>0$ and (iv) $(-,+)$ and $c<0$.

The full analysis and classification of these cases, which are
somewhat tortuous, are presented in the Appendix. The main results
for the viable cases are summarized in the Table \ref{TableSum}.
The opposite orbifold parities of $\Phi$ and $\Pt$ and the fact
that we can always define $\Phi$ to have even orbifold parity at
the IR brane permit the following simplification. \beq \bal
\label{srp}
 \textmd{V}_{\rm eff}
    =&\, U -\left[ \Pt_{\rm UV}F_{\rm UV}^+
- \Phi_{\rm UV}\Ft_{\rm UV}^+ +\Ft_{\rm
UV}\Phi_{\rm UV}^+ -F_{\rm UV} \Pt_{\rm UV}^+ \right]  + \hc\\
 &\,
 + \om^3 \left[ F_{\rm
IR}\Pt_{\rm IR}^- - \Phi_{\rm IR}\Ft_{\rm IR}^- \right] + \hc \eal
\eeq

\section{Supersymmetry Breaking}

 In this section, we consider the various supersymmetry
breaking contributions to the visible sector and calculate their
effects. There is a variety of ways by which high-energy
supersymmetry breaking is transmitted to us; 5-d gravity loops,
gauge mediation, direct mediation from unknown UV physics and
anomaly mediation. The 5-d gravity loop calculations is given by
\cite{Rattetal} and the soft masses induced by 1-loop gravity
effect are estimated to be, \beq m_{\rm soft, gravity}\sim \om^2 \eeq
where for estimation purposes, we do
not display the 1-loop factor but we do take it into account when
we compare relative strengths of SUSY breaking mechanisms. Also,
we will be using the normalization $M_5 =1$ and will do so for the
rest of the paper. $M_5$ can easily be restored by dimensional
analysis.

For the gauge mediation sector, we add a pair of vector-like
messenger fields $Q_M$ and $\bar{Q}_M$ of ${\bf 5}+{\bf 5^*}$
representation under the SM gauge group $SU(5)_{SM} \supset
SU(3)_c \times SU(2)_L \times U(1)_Y$ that will couple to $\Phi$
through the IR-brane-localized superpotential term, $\De W= \Phi
Q_M \bar{Q}_M$. From the point of view of the CFT, this
corresponds to composite bound states of the CFT that acquire
intermediate masses and convey supersymmetry breaking to our SM
fields through their $F$-terms.

The effective 4-d Lagrangian that characterizes the soft SUSY
breaking masses from the various mechanisms\footnote{With the
exception of gravity which we have estimated earlier.} can then be
parameterized as \beq \bal
    \scr{L}_{4d} &= -V_{\rm eff,\om} + \int d^4 \theta \, \om^{\dagger}\om \left[1 + (1+\Phi^{\dagger}_{\rm IR}\Phi_{IR})(Q^{\dagger}Q+ Q^{\dagger}_MQ_M+\bar{Q}^{\dagger}_M\bar{Q}_M) \right]\\
    &+ \int d^2\theta \, \beta\om^3\Phi \bar{Q}_M Q_M + \hc \eal \eeq
 where $Q$ denote SM chiral superfields and we have set most of the coefficients (except $\beta$, which though
also of order one, is essential for us to see that we have viable gauge mediation) to be
one for simplicity. The effective potential of the radion, $V_{\rm
eff, \om}$, can be found by redoing the entire calculation thus
far in terms of superfields and then through the method of spurion
analysis\cite{LS2}, promote the radius into a superfield and
finally extracting the requisite potential by singling out the
$F$-term of the radion, $F_{\om}$. We have only outlined the above
procedure as the cases of unique orbifold parities and $c$ lead us
to the same result as Ref. \cite{SwoS}. \beq[RadTadpole] \bal V_{\rm eff,\om}
= \om^2 \Biggl[ & 3 \left( -W + \sfrac 12 \Phi \frac{\d W}{\d
\Phi} \right)
\\
& + \sfrac 12 \om \frac{\d \Phi_{\rm IR}}{\d \om} \left( -\frac{\d
W}{\d \Phi} + \Phi \frac{\d^2 W}{\d \Phi^2} \right) \Biggr]_{\rm
IR} F_\om + \hc + \cdots \eal \eeq

  With the effective radion potential, we can determine
 the scale of anomaly mediation.\beq
m_{\rm soft, anomaly} \sim
\frac{F_{\om}}{\om}=\frac{1}{\om}\frac{\partial V_{\rm eff,\om}}
 {\partial \om} \sim  \om W \sim \om \Phi_{\rm IR}^n \sim
\LIR \om^{n\frac{d-5}{2n-3}}\eeq where $\LIR =\om M_5$ is the
cutoff of the 4-d effective theory localized on  the IR brane.

After canonical normalization, the direct mediation from the
Lagrangian above gives the following 
direct contribution to the SM chiral superfields. \beq
m_{\rm{soft, direct}}^2\sim F^{\dagger}_{\rm IR}F _{\rm IR}\eeq
This is generally flavor non-diagonal and causes SUSY FCNC
processes that are strongly constrained by experiments.

The hypermultiplet-messenger interaction term in the
superpotential completely fixes the parameters of our gauge
mediation sector. This allows us to write down the mass matrix of
the scalar messengers. \beq \label{scalmesmassmatrix}
    m_{\rm messenger}^2=\left(%
\begin{array}{cc}
  \beta^2 \om^{\dagger}\om |\Phi_{\rm IR}|^2 +  |F_{\rm IR}|^2 & \beta \om F_{\rm IR}  \\
 \beta \om^{\dagger} F_{\rm IR}^{\dagger} & \beta^2 \om^{\dagger}\om |\Phi_{\rm IR}|^2+ |F_{\rm IR}|^2  \\
\end{array}%
\right) \eeq Requiring that SM gauge symmetries (as the messenger
superfields are charged under the SM) not be broken gives us the
following constraint. \beq \beta^2 \om^{\dagger}\om |\Phi_{\rm
IR}|^2 + |F_{\rm IR}|^2 \geq |\beta \om F_{\rm IR}| \eeq
Obviously, the $|F_{\rm IR}|^2$ term should not dominate as
 we would otherwise have
$\LIR < |F_{\rm IR}|$. Taking this into account, the constraint
condition becomes $\beta \om |\Phi_{\rm IR}|^2 \geq |F_{\rm IR}|$.
The gauge mediation soft breaking masses are given by \beq
m_{\rm{soft, gauge}} \sim \frac{F_{\rm IR}}{\Phi_{\rm IR}}.\eeq

In order to have a viable gauge mediation scenario, we require
that $\Phi_{\rm IR} \ll \om^0$ as well, so that the messenger
scale is lower than $\LIR$ and that gauge mediation dominates over
direct mediation. As can be easily seen from the results given in
the Appendix, there are only two cases which satisfy the above
requirements: both orbifold parity assignments with $c<0$, $a_1
=0$ and $n \geq 3$. The solutions to the equations of motion for
these viable models are summarized in Table 1.

\begin{table}[t]
\caption{\label{TableSum}Summary of the results from the Appendix
for various brane quantities. We give the leading power $p$, where
$Q= \om^p +$ higher order term in $\om$, for two cases with $c<0$.
We have used $d \equiv |c|+\frac 52$, $\tilde{d} \equiv |c|+\frac
32$ and we parameterized the superpotential by $W=a_n\Phi^n$ plus
higher order terms with $n\geq 2$. As for the UV potential,  $U$
is taken to be, to leading order, a function of $\Phi_{\rm UV}$
for the $(+,+)$ case or a function of $\Pt_{\rm UV}^{q+1}$ for the
$(-,+)$ case,  $(q \geq 1)$.
}
\begin{center}
\begin{tabular}{|c|c|c|}
  \hline
Quantity & $(+,+)$ orbifold &$(-,+)$ orbifold \\
  (to leading order) &  parity for $\Phi$& parity for $\Phi$ \\
 \hline
  $F_{\rm IR}$ & $\frac{d-5}{2n-3}(n-1)+1$ &$\frac{d-5}{2n-3}(n-1)+1$\\
  $\Phi_{\rm IR}$ & $\frac{d-5}{2n-3}$ &$\frac{d-5}{2n-3}$\\
  $F_{\rm UV}$ & $\frac{d-5}{2n-3}(n-1)+d$ &$\frac{d-5}{2n-3}(n-1)+d$\\
  $\Phi_{\rm UV}$ & 0 &0\\
  $\Pt_{\rm UV}$ & $\frac{\partial U}{\partial F}$ &$(\frac{d-5}{2n-3}(n-1)+d)/q$\\
  $\Pt_{\rm IR}$ & $\frac{d-5}{2n-3}(n-1)$ &$\frac{d-5}{2n-3}(n-1)$\\
  $\Ft_{\rm UV}$ & 0 &0\\
  $\Ft_{\rm IR}$ & $d-4$ &$d-4$\\
  \hline
\end{tabular}
\end{center}
\end{table}

In a nutshell, the soft contributions from different mechanisms
for transmission of SUSY breaking are, to leading order, given
below.
 \beq
    m_{\rm soft} \sim \left \{ \begin{array}{cll}
     \frac{F_{\rm IR}}{\Phi_{\rm IR}} &\sim \LIR \om^{\frac{d-5}{2n-3}(n-2)}& gauge \\
      F_{\rm IR} &\sim \LIR \om^{\frac{d-5}{2n-3}(n-1)}&direct \\
    \frac{F_{\om}}{\om} &\sim\Lambda_{IR}\om^{\frac{d-5}{2n-3}n}&anomaly\\
    &\sim \Lambda_{IR}\om &gravity
    \end{array}\right .
\eeq
We see that for $n \geq 3$ and $c < - \frac{5}{2}$ ($d > 5$), gauge mediation
always dominates over direct and anomaly mediation and
will dominate over gravity in phenomenologically viable regions.

%
%
%

The radion mass can be found from the scalar potential given in
eq.(\ref{srp}) and the sub-dominant contribution is given by
$F_{\om}^{\dagger}F_{\om}$. The radion mass from the latter can be
estimated quickly as $\Delta m_{\rm radion} \sim
\frac{F_{\om}}{\om}$ which is of order $m_{\rm soft, anomaly}$.
For the other contributions, as we need all the possible
$\om$-dependent leading powers of $\om$, knowing that the leading
order of $\Phi_{UV}$ and $\Ft_{UV}$ are $\scr{O}(1)$ in Planckian
units may not be enough. It is possible that the next order, which
is the leading $\om$-dependent order, may contribute. Although it
turns out to be unimportant, we give the result here for
completeness.\beq
    \Ft_{\rm UV}\sim \Phi_{\rm UV} \sim 1+\Phi_{\rm IR}\om^{\tilde{d}}
\eeq The potential is dominated by the first term in
eq.(\ref{srp}) and we found \beq
    V \sim \om^4\om^{2(n-1)\frac{d-5}{2n-3}}
\eeq This gives us the radion mass,\beq
    m_{\rm radion} \sim \LIR\om^{(n-1)\frac{d-5}{2n-3}}
\eeq which is the same order as the direct mediation soft masses.
In our case, the radion mass is around 10 GeV. Due to the
effective 4-d theory cutoff ~TeV$\ll\Lambda_{\rm IR}\ll M_{pl}$,
a 10 GeV radion is not ruled out by either collider search or
cosmological observation. We will discuss this in more detail in
the next section. If the same potential is to be used to stabilize
the radius, it has to be at least bigger than the potential due to
the Casimir effect\footnote{An interesting possibility for radius
stabilization is through an interplay of bulk hypermultiplet and
Casimir effects.} which is estimated to be $\om^6$ \cite{Pometal}.
From the results given above, we see that $d$ should be $d \leq
7-1/(n-1)$.

\section{Stabilization and Phenomenology}

We now analyze a specific model and discuss the stabilization and
phenomenological issues involved. We pick the $(+,+)$ orbifold
parity condition as well as $d > 5$ (which also means $c<
-\sfrac52$) and choose a potential on the UV brane and a
superpotential on the IR brane of the following form. \beq
    U = b (\Phi_{\rm UV}+\Phi_{\rm UV}^\dagger)\,,\hspace{0.5in}
    W = a \Phi_{\rm IR}^3  \eeq

We discard the higher order terms of $\Phi_{UV}$ in $U$ for
simplicity. Restoring these terms does not affect the results
significantly as our equations contain only first derivatives in
$\Phi_{UV}$ and since $b$ is of order one. Although $F$ dependent
term can be added to $U$, this modification will only change the
result of $\Ft_{\rm UV}$ which is not relevant for our purpose.

Solving the equations of motion yields the following relations.
\beq \bal
    \Phi_{\rm IR} &= -\left(\frac{b}{(3a)^2(2d-4)k}\right)^{\frac{1}{3}}\om^{\frac{d-5}{3}} + ...\\
    F_{IR} &= -\left(\frac{b^2 k(d-2)}{12 a}\right)^\frac{1}{3} \om^{\frac{2d-7}{3}} + ...\\
    \Phi_{\rm UV} &= -\frac{b}{2(2 d -4)k}+\Phi_{\rm
    IR}\om^{\tilde{d}} \eal
\eeq The jump conditions across the brane then determines the
solution of the other fields. \beq \bal
    \Ft_{\rm UV} &= -\frac{1}{2}b\\
    \tilde{\Phi}_{\rm IR} &= -\frac{1}{2}\left(\frac{b^2}{3ak^2 (2d-4)^2 }\right)^{\frac{1}{3}}\om^{2\frac{d-5}{3}}+ ...\\
    \tilde{\Phi}_{\rm UV} &= 0
\eal \eeq



The final form of the effective potential, to leading order, can
then be obtained. \beq[Veffrace1]
    V_{\rm eff}  =\frac{3b}{4}\Phi_{\rm IR}\om^{d-1} + \ldots
    = -\frac{1}{2}\left(\frac{3 b^4}{2 a^2 k (d-2)}\right)^{\frac{1}{3}}\om^{4\frac{d-2}{3}}
    +\ldots
\eeq Notice that the sign of the potential is always negative. In
order to have a racetrack-type stabilization, we need to include
another contribution to the effective potential which would give a
leading order term with a positive sign. To do that, let us
introduce another bulk hypermultiplet, $\Psi$ (which corresponds to a CFT
operator dimension $d'$), with the same $(+,+)$ orbifold parity conditions
but we flip $c$'s sign\footnote{Although the $(+,+)$ orbifold
boundary
condition with $c > 0 $ does not lead to viable
gauge mediation, it can however be
used for stabilization.} so that $c > 0$. The brane-localized
potential and superpotential for $\Psi$ is,
\beq U = b' \Psi_{\rm UV} ^2 + b'_2 F +\hc \,,\hspace{0.5in}
    W =  a' \Psi_{\rm IR}^2
\eeq which leads to an effective potential due to $\Psi$ of the form
\beq[Veffrace2] V_{\rm eff,\Psi} =
\left[1-  \frac{a'^2(d'-3)}{d'-2}\right]b'^2 _2 (d'-3)k
\om^{2d'-6} +\ldots
\eeq where for small enough $a'$ we can get a positive contribution to the effective potential.
For racetrack stabilization to occur, we can compare the $\om$-dependence of Eq.(\ref{eq:Veffrace1}) with the
above and conclude that \beq[dpcond] d'= \frac{2d + 5}{3} + \ep \eeq where $\ep$ is an $\scr{O}(\frac{1}{10})$
positive number. Now, we need to check the SUSY breaking to ensure our earlier discussion is not
invalidated by the presence of this hypermultiplet. First, the direct mediation contribution is
parameterized by \beq F_{\rm IR, \Psi} = b'_2 (d'-3)k\om^{d'-4} + \ldots \sim F_{\rm IR, \Phi} \om^\ep\eeq
on substitution of Eq.(\ref{eq:dpcond}) and so we have a subdominant contribution to FCNCs from
this hypermultiplet. As for the anomaly-mediated contribution,
plugging in our IR superpotential into Eq.(\ref{eq:RadTadpole}),
we obtain a vanishing coefficient for the $F_\om$ tadpole term which is to be expected as this term
preserves conformal symmetry ($a'$ is dimensionless). The gravity loop contributions are the same while
we do not need to consider gauge mediation from this sector as we are not coupling $\Psi$ to the messengers.
It is clear that the effects of the $\Psi$ field are subdominant and for the
rest of the paper, we will only consider the contributions
from the $\Phi$ field.

Having the full effective potential would allow us to work out the
radion mass in this scenario, \beq
    m^2_{\rm radion}\sim\frac{20(d'-d)}{9}
    \left(\frac{b^4(d-2)^2}{18a^2k}\right)^{\frac{1}{3}}\om^{4\frac{d-5}{3}}\LIR^2
\eeq where we have expressed it in terms of $\LIR = \om M_5$. To obtain actual values,
one must note the coefficient in front of $\LIR$ is still expressed in units where $M_5 =1$.

We are now ready to discuss the phenomenology of our model.
Assuming a low-energy MSSM content, the gauge mediation
contribution to the sparticle masses is roughly given by (for
$n=3$ case) \cite{GMSBreview} \beq
  m_{\rm soft} \sim \frac{\alpha_{\rm SM}}{4 \pi}
   \frac{F_{\rm IR}}{\Phi_{\rm IR}}
   \sim 10^{-2} \LIR  \om^{\frac{d-5}{3}},
\label{GMSBsoftmass} \eeq where $\alpha_{\rm SM}$ stands for the
SM gauge coupling constants. Considering that the messenger scale
is given by $M_{\rm mess} \sim \om \Phi_{\rm IR} \sim \LIR
\om^{\frac{d-5}{3}}$, we find $M_{\rm mess} \sim 10^2 m_{\rm soft}
\sim 10-100$ TeV for the natural scale of the sparticle mass
$m_{\rm soft}=100-1000$ GeV, independent of the parameters of the
model.

We also have the direct mediation contribution, \beq
  m_{\rm{direct}}
  \sim  F_{\rm IR}
     \sim \LIR  \om^{2\frac{d-5}{3}}.
\eeq This is flavor-dependent and should be a sub-dominant
contribution compared to the gauge mediation contribution being
flavor blind. Define the ratio as \beq
 \epsilon = \frac{m_{\rm{direct}}}{m_{\rm soft}}
         \sim 10^2 \om^{\frac{d-5}{3}}.
\eeq Using this and the relation $\LIR  = M_5 \omega$,
Eq.(\ref{GMSBsoftmass}) leads to the relation between $d$ and
$\epsilon$, \beq d = 5+ 3 \frac{\log( 10^{-2} \epsilon) } {\log
\left( \frac{10^4 m_{\rm soft}}{\epsilon M_5} \right)}. \eeq FCNC
processes induced by flavor dependent soft terms are strongly
constrained by experiments, roughly $\epsilon \leq 10^{-2}$
\cite{SUSYFCNC}. This gives the lower bond on $d \geq 6.16$ when
we take $m_{\rm soft}=100$ GeV and $M_5 = 2.4 \times 10^{18}$ GeV
(reduced Planck mass). Recall that there exists an upper bound on
$d \leq 6.5$ as discussed in the previous section. Therefore, the
parameter $d$ (in the $n=3$ case) should lie in the range $6.16
\leq d \leq 6.5$ which corresponds to $ 1.6 \times 10^{-3} \leq
\epsilon \leq 10^{-2}$ and a compositeness scale of $ 6.2 \times
10^8 \mathrm{GeV} \geq \LIR \geq 1.0 \times 10^8 \mathrm{GeV}$. It
is extremely interesting that the upper bound on $d$ gives the
lower bound on $\epsilon$ being an order of magnitude below
current experimental bounds. We may expect that future experiments
will reveal a sizable FCNC originating from flavor-dependent soft
masses. In comparison, conventional gauge mediation models have
negligibly small FCNC predictions.

There is another crucial difference between our model from
conventional models. The gravitino and cosmological moduli, by
virtue of SUSY being an accidental symmetry, have Planckian masses
and therefore completely decouple from the low energy
phenomenology. This is in contrast to the usual gauge mediation
models where gravitino is always the LSP. The feature where
gravitino is not the LSP in a gauge mediation model was first
proposed in Ref.~\cite{Nomura-Yanagida} in a different context. In
their model, the gravitino mass lies in the range of 100 GeV to 1
TeV and consequently still suffers from the gravitino and
cosmological moduli problems \cite{gravitino-problem1}
\cite{gravitino-problem2}. In our model, there is no gravitino
problem, because the superheavy gravitino cannot be produced in
the early universe.

Assuming R-parity conservation, the LSP neutralino is the most
reasonable candidate for dark matter. This case was first
investigated in detail in Ref.~\cite{Nomura-Suzuki} and it was
found that in a wide range of parameter space, the neutralino LSP
is primarily composed of the B-ino and it can constitute the
dominant component of the dark matter through co-annihilation
processes with the right-handed scalar leptons. Note that recent
cosmological observations, especially the Wilkinson Microwave
Anisotropy Probe (WMAP) satellite~\cite{WMAP}, have established
the relic density of the cold dark matter with great accuracy (in
the $2\sigma$ range),
\begin{eqnarray}
\Omega_{\mathrm{CDM}} h^2 = 0.1126^{+0.0161}_{-0.0181}. \nonumber
\end{eqnarray}
In addition, recent results of LEP-2 have pushed up the lower
bound on the lightest Higgs mass, $m_h \geq 114$ GeV. These recent
results will dramatically reduce the allowed parameter region
previously obtained in \cite{Nomura-Suzuki}, and updating the
previous results would be a relevant and worthwhile exercise. We
will give a full detailed phenomenological studies including
additional experimental considerations as well as the latest
cosmological and astrophysical observations in a forthcoming paper
\cite{GNO2}.

Before concluding, we briefly consider phenomenology and cosmology
related to the radion. One might naively be tempted to conclude
that we have exchanged the problems associated with the gravitino
in conventional gauge mediation for ones associated with the
radion in the present scenario. But actually, the radion behaves
in a very different way from the gravitino. Although its precise
value depends on parameters in the model including those in the
brane potentials, the mass scale of the radion lies around 10 GeV.
After electroweak symmetry breaking, through mixing with neutral
Higgs boson, the radion couples to the SM particles with strength
of $\sim y\frac{v}{\Lambda_{\rm IR}}$ \cite{radioncosmo}, where
$y$ is the Yukawa coupling constant and $v$ is the Higgs vev. This
coupling is very much suppressed and so the radion totally
decouples from the collider phenomenology.

The most stringent constraint actually comes from cosmological
considerations. There is a possibility that the coherent
oscillation of the radion to dominate the energy density of the
early universe at a low temperature and its decay into SM
particles to reheat the universe. In order not to change the
successful predictions of Big Bang Nucleosynthesis (BBN), the
reheating temperature should exceed the temperature of the BBN
era, typically $\cal{O}$(1 MeV). For a radion mass of around 10
GeV, the radion decays mainly into $b\bar{b}$ and $\tau
\bar{\tau}$ and its decay width can be estimated as\beq
 \Gamma \sim \frac{m_{\mathrm{radion}} m_b^2}{\Lambda_{\rm IR}^2}
 \sim 10^{-15} \mbox{GeV} ,
\eeq which gives us a reheating temperature of the order \beq
 T_{RH} \sim \sqrt{\Gamma M_{\mathrm{Pl}}}\sim  50 \mbox{GeV},
\eeq which is high enough not to affect BBN. Note that this
tempetature is also sufficient for a neutralino dark matter with
mass around 100 GeV (which we have discussed earlier) to be in
thermal equilibrium.

\section{Conclusions}
We have considered various five-dimensional brane and bulk
configurations to determine generic setups that would allow the
implementation of the ``Supersymmetry without Supersymmetry''
paradigm in a gauge mediation setup as opposed to the traditional
dynamical supersymmetry breaking. Requiring that the dominant
contribution to the sparticle masses be through gauge mediation,
instead of anomaly, direct or gravity mediation, and that flavor
changing contributions from direct mediation be lower than current
experimental bounds, we have established a region of parameter
space whereby a class of these gauge mediation from emergent
supersymmetry theories can naturally exist. In addition to having
no small parameters in our theory, the parameters of the gauge
mediation sector are completely fixed by the conformal dynamics.

We find that the racetrack stabilization of the extra dimension,
or alternatively the spontaneous breaking of the conformal
symmetry through irrelevant operators from the CFT viewpoint,
necessarily dictates that the FCNCs are only suppressed to the
extent that it is, at the minimum, an order magnitude below
current experimental bounds. This is very different from
conventional GMSB models where the FCNC contributions are
negligibly small. This setup not only solves the hierarchy problem
and the SUSY flavor problem but also averts the gravitino problem
by completely decoupling it from low-scale physics as it receives
mass corrections of Planckian order. The radion is the only
low-energy degree of freedom in this theory besides the SM fields
and their supersymmetric partners. But it decouples from collider
phenomenology and does not affect Big Bang Nucleosynthesis.
Assuming R-parity conservation, the neutralino LSP (dominantly
B-ino) from this class of gauge-mediated theories can also provide
a cold dark matter candidate.

 This class of models frees up large regions of
parameter space for gauge mediation that were previously excluded
in the conventional picture. Many interesting model-building
possibilities and directions await further exploration with this
explicit realization of gauge mediation from emergent
supersymmetry.
\newpage

\section*{Acknowledgements}
The work of H.S.G., S.P.N. and N.O. were respectively supported by
the NSF (under grant PHY-0408954), the DOE (under contract
DE-FG02-91ER40626) and the Grant-in-Aid for Scientific Research in
Japan (\#15740164). The authors would like to thank Z. Chacko, I.
Gogoladze,  A. Iglesias, Jason Lee, M.A. Luty and  Q. Shafi  for
useful discussions. We would also like to record our gratitude to
Z. Chacko for his invaluable comments on the final manuscript.
S.P.N. also wishes to thank R. Kitano, G. Kribs, Tianjun Li and A.
Sirlin for sitting through his Russian-length seminar at the IAS.
N.O. would like to thank the Theoretical Elementary Particle
 Physics Group of the University of Arizona, and especially Z. Chacko,
 for their hospitality at the beginning of this work. He would also like to thank the Elementary Particle Physics Group
 at the University of Maryland, and especially M.A. Luty
 for their hospitality during the completion of this work.

\startappendices
\section*{Appendix: Classification of General Solutions}
In this section, we classify the general solutions that arise
after the imposition of the orbifold parity and the selection of
the sign of $c$. Before we go any further, we would like to
introduce some notational changes that will help us understand the
underlying conformal dynamics of this model. To this end, we can
use the fact that we have large anomalous dimensions to re-express
the dimensions of the operators corresponding to $\Phi$ and $\Pt$
respectively as \beq
    d \equiv |c|+\frac 52  \,,\hspace{0.5in}
    \tilde{d} \equiv |c|+\frac 32
\eeq and we parameterize the superpotential by
$W=a_1\Phi+a_n\Phi^n$ with $n\geq 2$.

\subsection{\textbf{Case (i): $(+,+)$ and $c>0$}}

The equations that have to  be solved are
\begin{eqnarray}
    \frac{\d U}{\d \Phi_{\rm UV}}&=&\frac{\d^2 W}{\d \Phi_{\rm IR}^2} F_{\rm UV} \om^{(d+\tilde{d} - 4)}\\
    \frac{\d U}{\d F_{\rm UV}}&=&\frac{\d W}{\d \Phi_{\rm IR}} \om^{\tilde{d}}  + 2 \frac{F_{\rm
UV}^\dagger(1-\om ^{2\tilde{d}-4})}{(2 \tilde{d} - 4) k} \\
    \Phi_{\rm IR} &=& \Phi_{\rm UV}  \om^{(d-4)} +
\frac{F_{\rm UV} ^\dagger}{2(2d - 4) k} (\frac{\d^2 W}{\d
\Phi_{\rm IR}^2}) ^\dagger \om^{(d - 5)} (1-\om ^{2d-4})
\eql{eq1a3}
\end{eqnarray}
The other parameters can be obtained from
\begin{eqnarray}
    \Pt_{\rm IR} &=& -\frac{1}{2}\frac{\d W}{\d \Phi_{\rm IR}}\\
    \Ft_{\rm UV} &=& \Ft_{\rm IR} \om^{\tilde{d}}=-\frac{1}{2} \frac{\d U}{\d
\Phi_{\rm UV}}\\
    F_{\rm IR} &=& F_{\rm UV} \om^{d - 4}\\
    \Pt_{\rm UV} &=& -\frac{1}{2}\frac{\d U}{\d F_{\rm UV}}
\end{eqnarray}

To estimate the size of these fields, we note that for a generic UV
potential, both $\Phi_{\rm UV}$ and $F_{\rm UV}$ are of
$\scr{O}(1)$ and satisfy,
\begin{eqnarray}
    \frac{\d U}{\d \Phi_{\rm UV}}&=&0\nonumber\\
    \frac{\d U}{\d F_{\rm UV}}&=&0
\end{eqnarray}
Although we can have a model with suppressed $\Phi_{\rm UV}$ or
$F_{\rm UV}$, or both, we restrict ourselves to a class of model
where $F_{\rm UV}$ is of order one as $F$ is then UV dominated. A small
$\Phi_{\rm UV}$ does not provide interesting models for gauge
mediation by itself since $\Phi_{\rm IR}$ is smaller in this case
and so would be harder to meet the requirements of gauge mediation
model-building. As we will see below, even $\Phi_{\rm UV} \sim 1$
is insufficient for gauge mediation.

With $F_{\rm UV}$ and $\Phi_{\rm UV}$ set to one, \Eq{eq1a3} gives
us,
\begin{eqnarray}
    \Phi_{\rm IR} &\sim&  \left \{ \begin{array}{ccl}
      \om ^{\tilde{d}-4} && ,a_2 \neq 0 \\
      \om ^{d-4} && ,a_2 = 0 \\
    \end{array} \right .\\
    F_{\rm IR} &\sim&  \om ^{d-4}
\end{eqnarray}

\subsection{\textbf{Case (ii): $(+,+)$ and $c<0$}}

The equations that have to  be solved are \beq[eq2a1]
    \frac{\d U}{\d \Phi_{\rm UV}}\om ^{d-4}&=\frac{\d^2 W}{\d \Phi_{\rm IR}^2} F_{\rm IR} \\
    \eql{eq2a2}
    \frac{\d U}{\d F_{\rm UV}}\om^{d-4}&=\frac{\d W}{\d \Phi_{\rm IR}} + 2 \frac{F_{\rm
IR}^\dagger(1-\om ^{2d-4})}{(2 d - 4) k}\om^{-1} \\
\eql{eq2a3}
    \Phi_{\rm UV} \om^{d-4}&= \Phi_{\rm IR}  \om^{d+\tilde{d}-4} -
\frac{F_{\rm IR} ^\dagger}{2(2\tilde{d} - 4) k} (\frac{\d^2 W}{\d
\Phi_{\rm IR}^2}) ^\dagger  (1-\om ^{2\tilde{d}-4}) \eeq The other
parameters are then obtained from \beq[eq2b1]
    \Pt_{\rm IR} &= -\frac{1}{2}\frac{\d W}{\d \Phi_{\rm IR}}\\
\eql{eq2b2}
    \Ft_{\rm UV} &= \Ft_{\rm IR} \om^{4-d}=-\frac{1}{2} \frac{\d U}{\d
\Phi_{\rm UV}}\\
\eql{eq2b3}
    F_{\rm IR} &= F_{\rm UV} \om^{-\tilde{d}}\\
\eql{eq2b4}
    \Pt_{\rm UV} &= -\frac{1}{2}\frac{\d U}{\d F_{\rm UV}}
\eeq In this case, $\Ft_{\rm UV}$ is UV dominated and so we
consider only $\Ft_{\rm UV}=\scr{O}(1)$ which, from \Eq{eq2b2} and
the fact that $F$ is IR dominated, requires $U= U_1(\Phi_{\rm
UV})+\ldots $. From \Eqs{eq2a1} and \eq{eq2a3}, $\Phi_{\rm UV}$ is
also $\scr{O}(1)$. Note that it seems we have to use a special UV
potential in order to achieve our goal but all it requires is
actually that there be at least one term in the potential which is
$F$-independent. As $F_{\rm IR}$ is expected to be small, extra
$F$-dependent terms will not change our conclusions and different
$\Phi$-dependent terms will only change the minor details of the
result without affecting our main conclusions. The equations we
have to solve are then reduced to \beq[eq2a4]
    \frac{\d U}{\d F_{\rm UV}}\om^{d-4}&\sim\frac{\d W}{\d \Phi_{\rm IR}} + 2F_{\rm IR}^\dagger\om^{-1} \\
\eql{eq2a5}
    F_{\rm IR} \frac{\d^2 W}{\d\Phi_{\rm IR}^2} &\sim \om^{d-4} \eeq

For $a_1 \neq 0$, $\frac{\d W}{\d \Phi}\sim \scr{O} (1)$. Hence
from \Eq{eq2a4}, $F_{\rm IR}\sim \om$. It is then implied by
\Eq{eq2a5} that $\frac{\d^2 W}{\d \Phi^2} \sim \om^{d-5}$ and so
$\Phi_{\rm IR} \sim \om^{\frac{d-5}{n-2}}$ for $n>2$ (this does
not lead to viable gauge mediation). For the special case where
$n=2$ it is easy to see that the only solution is $\Phi_{\rm IR}
\sim \scr{O} (1)$ and $F_{\rm IR}\sim \om^{d-4}$ (this again does
not lead to viable gauge mediation as the messenger scale would be at
Planck scale).

For $a_1 =0 $, the equations above can be solved by observing from
\Eq{eq2a5} that $F_{\rm IR}\geq \om^{d-4}$. So we can simply set
the left-hand side of \Eq{eq2a4} to zero. The solutions are then
found to be,
\begin{eqnarray}
    F_{\rm IR} &=& \om^{\frac{d-5}{2n-3}(n-1)+1}\\
    \Phi_{\rm IR} &=& \om^{\frac{d-5}{2n-3}}
\end{eqnarray}

\subsection{\textbf{Case (iii): $(-,+)$ and $c>0$}}

We consider the case where $F_{\rm UV} \sim \scr{O}(1)$. This
implies $\Pt_{\rm UV} \sim \scr{O}(1)$ and $\frac{\d U}{\d
\Pt_{\rm UV}}=f(\Pt_{\rm UV})+ \ldots$ .

The equations that we need to solve are
 \beq[eq3a1]
    \Phi_{\rm IR} &\sim (\frac{\d U}{\d \Ft_{\rm UV}}+\frac{\d^2 W}{\d \Phi_{\rm
    IR}^2}\om^{-1})\om^{d-4}\\
\eql{eq3a2}
    \Ft_{\rm UV}  &\sim \frac{\d^2 W}{\d \Phi_{\rm IR}^2} \om^{d+\tilde{d}-4}
\eeq
The other parameters can be obtained from \beq[eq3b1]
   \Pt_{\rm IR} &= -\frac{1}{2}\frac{\d W}{\d \Phi_{\rm IR}}\\
\eql{eq3b2}
    F_{\rm IR}&=  F_{\rm UV} \om^{d-4}\\
\eql{eq3b3}
    \Ft_{\rm IR} &= \Ft_{\rm UV} \om^{-\tilde{d}}\\
\eql{eq3b4}
    \Phi_{\rm UV} &=  \frac{1}{2}\frac{\d U}{\d \Ft_{\rm UV}}
\eeq For $a_2\neq 0$, the second term on the RHS of \eq{eq3a1}
dominates and so we have $\Phi_{\rm IR}\sim \om^{d-5}$. For
$a_2=0$, this term is always less than $\Phi_{\rm IR}$ and so can
be discarded. The solution of $\Phi_{\rm IR}$ is determined from
the $\frac{\d U}{\d \Ft_{\rm UV}}$ term. If $U$ has a term linear
in $\Ft$, $\frac{\d U}{\d \Ft_{\rm UV}}\sim \scr{O} (1)$ and so
$\Phi_{\rm IR}\sim \om^{d-4}$. Otherwise from \eq{eq3a2},
$\frac{\d U}{\d \Ft_{\rm UV}}$ has to be less than $\Phi_{\rm IR}$
as well. In that case, $\Phi_{\rm IR}$ has only a trivial
solution. We summarize the results for this case in the following,
\begin{eqnarray}
    \Phi_{\rm IR} &=& \left \{ \begin{array}{cccl}
      \om^{d-5} && ,a_2\neq 0 &\\
      \om^{d-4} && ,a_2 =0 & ,\frac{\d U}{\d \Ft_{\rm UV}}\sim\scr{O} (1)\\
      0         && ,a_2 =0 & ,\frac{\d U}{\d \Ft_{\rm UV}}\sim \mathrm{otherwise}\\
    \end{array}\right . \\
    F_{\rm IR} &=& \om^{d-4}
\end{eqnarray}
which are the same as that obtained in case (i).
\subsection{\textbf{Case (iv): $(-,+)$ and $c<0$}}

Again, we consider only the case where $\Ft_{\rm UV} \sim 1$. This
implies that $\Phi_{\rm UV}\sim 1 $ and $\frac{\d U}{\d \Ft_{\rm
UV}}= f(\Ft_{\rm UV})+\ldots$.

The equations that we have to solve are
 \beq[eq4a1]
    F_{\rm IR} \om^{\tilde{d}} &\sim  \frac{\d U}{\d \Pt_{\rm UV}}\\
\eql{eq4a2}
    \frac{\d^2 W}{\d \Phi_{\rm IR}^2}F_{\rm IR} &= \om^{d-4}\\
\eql{eq4a3}
    \frac{\d W}{\d \Phi_{\rm IR}}  &= \Pt_{\rm UV}\om^{d-4}+F_{\rm IR}\om^{-1}
\eeq

The potential $U$  can be parameterized by $\frac{\d U}{\d\Pt}\sim
\Pt^{q}$ where $q$ is some integer. In order to have a solution
with $F_{\rm IR}\ll 1$, we have to choose a potential $U$ with
$q\neq 0$. We can also simplify \Eq{eq4a3} by keeping only the
dominant term on the RHS of the equation. The second term always
dominates as we can see from \Eq{eq4a2} that $F_{\rm IR} \geq
\om^{d-4}$.

For a generic potential, there is always the possibility that a
solution with $\Phi_{\rm IR}\sim 1$. Hence from \Eq{eq4a2},
$F_{\rm IR} \sim \om^{d-4}$ exist. These solutions are not
interesting and require higher order terms in the  potential. We
will ignore these solutions and concentrate on those with small
$\Phi_{\rm IR}$. Let us look at the case where both $a_1$ and
$a_2$ are not vanishing. It is obvious from \Eqs{eq4a2} and
\eq{eq4a3} that these potential falls into the category above. If
$a_1\neq 0$ and $a_2 =0$ (e.g. $n>2$), \Eq{eq4a3} implies that
$F_{\rm IR}\sim \om$ and \Eq{eq4a2} implies $\Phi_{\rm
IR}\sim\om^{\frac{d-5}{n-2}}$. For the case $a_1=0$, \Eq{eq4a3}
becomes
\beq[eq253]
   \Phi_{\rm IR}^{n-1}  &= F_{\rm IR}\om^{-1}
\eeq
The solution is then found to be \beq
    \Phi_{\rm IR} &= \left \{ \begin{array}{ccl}
      \om^{\frac {d-5}{n-2}} && ,a_1\neq 0, a_2=0\\
      \om^{\frac{\tilde{d}-4}{2n-3}} && ,a_1 =0 \\
    \end{array}\right .\\
    F_{\rm IR} &= \left \{ \begin{array}{ccl}
      \om && ,a_1\neq 0, a_2=0\\
      \om^{\frac{\tilde{d}(n-1)-(2n-1)}{2n-3}} && ,a_1 =0 \\
    \end{array}\right .
\eeq


\begin{thebibliography}{}

\bibitem{GMSB}
M.~Dine, W.~Fischler and M.~Srednicki,
Nucl.\ Phys.\ B {\bf 189}, 575 (1981); S.~Dimopoulos and S.~Raby,
Nucl.\ Phys.\ B {\bf 192}, 353 (1981); L.~Alvarez-Gaume,
M.~Claudson and M.~B.~Wise,
Nucl.\ Phys.\ B {\bf 207}, 96 (1982); M.~Dine and A.E.~Nelson,
Phys.\ Rev.\ D {\bf 48}, 1277 (1993) [hep-ph/9303230]; M.~Dine,
A.E.~Nelson and Y.~Shirman,
Phys.\ Rev.\ D {\bf 51}, 1362 (1995) [hep-ph/9408384]; M.~Dine,
A.E.~Nelson, Y.~Nir and Y.~Shirman,
Phys.\ Rev.\ D {\bf 53}, 2658 (1996) [hep-ph/9507378];
H.~Murayama,
Phys.\ Rev.\ Lett.\  {\bf 79}, 18 (1997) [hep-ph/9705271];
S.~Dimopoulos, G.R.~Dvali, R.~Rattazzi and G.F.~Giudice,
Nucl.\ Phys.\ B {\bf 510}, 12 (1998) [hep-ph/9705307]; M.A.~Luty,
Phys.\ Lett.\ B {\bf 414}, 71 (1997) [hep-ph/9706554].

\bibitem{dynSUSY}
K.A.~Intriligator, N.~Seiberg and S.H.~Shenker,
Phys.\ Lett.\ B {\bf 342}, 152 (1995) [hep-ph/9410203];
H.~Murayama,
Phys.\ Lett.\ B {\bf 355}, 187 (1995) [hep-th/9505082]; E.~Poppitz
and S.P.~Trivedi,
Phys.\ Lett.\ B {\bf 365}, 125 (1996) [hep-th/9507169]; K.I.~Izawa
and T.~Yanagida,
Prog.\ Theor.\ Phys.\  {\bf 95}, 829 (1996) [hep-th/9602180];
K.A.~Intriligator and S.~Thomas,
Nucl.\ Phys.\ B {\bf 473}, 121 (1996) [hep-th/9603158]. For a
review, see E.~Poppitz and S.~P.~Trivedi,
  Ann.\ Rev.\ Nucl.\ Part.\ Sci.\  {\bf 48}, 307 (1998)
  [arXiv:hep-th/9803107] or Y.~Shadmi and Y.~Shirman,
  Rev.\ Mod.\ Phys.\  {\bf 72}, 25 (2000)
  [arXiv:hep-th/9907225].

\bibitem{MoroiYanagida}
  T.~Asaka, J.~Hashiba, M.~Kawasaki and T.~Yanagida,
  Phys.\ Rev.\ D {\bf 58}, 083509 (1998)
  [arXiv:hep-ph/9711501];
  T.~Moroi,
  Phys.\ Rev.\ D {\bf 58}, 124008 (1998)
  [arXiv:hep-ph/9807265].

\bibitem{SwoS}
  H.~S.~Goh, M.~A.~Luty and S.~P.~Ng,
  JHEP {\bf 0501}, 040 (2005)
  [arXiv:hep-th/0309103].

\bibitem{GherghettaPom}
  T.~Gherghetta and A.~Pomarol,
  Phys.\ Rev.\ D {\bf 67}, 085018 (2003)
  [arXiv:hep-ph/0302001].

\bibitem{Kaplan}
  D.~B.~Kaplan,
  Phys.\ Lett.\ B {\bf 136}, 162 (1984).

\bibitem{Noscale}
    M.~A.~Luty and N.~Okada,
  JHEP {\bf 0304}, 050 (2003)
  [arXiv:hep-th/0209178].

\bibitem{AdSCFT}
E.~Witten,
  Adv.\ Theor.\ Math.\ Phys.\  {\bf 2}, 253 (1998)
  [arXiv:hep-th/9802150];
 S.~S.~Gubser, I.~R.~Klebanov and A.~M.~Polyakov,
  Phys.\ Lett.\ B {\bf 428}, 105 (1998)
  [arXiv:hep-th/9802109].

\bibitem{ConfSeq}
 L.~Randall and R.~Sundrum,
  Nucl.\ Phys.\ B {\bf 557}, 79 (1999)
  [arXiv:hep-th/9810155];  M.~A.~Luty and R.~Sundrum,
  Phys.\ Rev.\ D {\bf 65}, 066004 (2002)
  [arXiv:hep-th/0105137].

\bibitem{Strassler}
  M.~J.~Strassler,
  arXiv:hep-th/0309122.

\bibitem{RS1}
  L.~Randall and R.~Sundrum,
  Phys.\ Rev.\ Lett.\  {\bf 83}, 3370 (1999)
  [arXiv:hep-ph/9905221].

\bibitem{MartiPom}
  D.~Marti and A.~Pomarol,
  Phys.\ Rev.\ D {\bf 64}, 105025 (2001)
  [arXiv:hep-th/0106256].

\bibitem{Rattetal}
  T.~Gregoire, R.~Rattazzi, C.~A.~Scrucca, A.~Strumia and E.~Trincherini,
  Nucl.\ Phys.\ B {\bf 720}, 3 (2005)
  [arXiv:hep-th/0411216].
  For the flat case, see  R.~Rattazzi, C.~A.~Scrucca and A.~Strumia,
  Nucl.\ Phys.\ B {\bf 674}, 171 (2003)
  [arXiv:hep-th/0305184] as well as I.~L.~Buchbinder, S.~J.~J.~Gates, H.~S.~Goh, W.~D.~.~Linch, M.~A.~Luty, S.~P.~Ng and J.~Phillips,
  Phys.\ Rev.\ D {\bf 70}, 025008 (2004)
  [arXiv:hep-th/0305169].

\bibitem{LS2}
  M.~A.~Luty and R.~Sundrum,
  Phys.\ Rev.\ D {\bf 64}, 065012 (2001)
  [arXiv:hep-th/0012158].
  For the flat case, see M.~A.~Luty and R.~Sundrum,
  Phys.\ Rev.\ D {\bf 62}, 035008 (2000)
  [arXiv:hep-th/9910202].

\bibitem{Pometal}
  J.~Garriga and A.~Pomarol,
  Phys.\ Lett.\ B {\bf 560}, 91 (2003)
  [arXiv:hep-th/0212227].

\bibitem{GMSBreview}
 For a review, see
 G.~F.~Giudice and R.~Rattazzi,
  Phys.\ Rept.\  {\bf 322}, 419 (1999)
  [arXiv:hep-ph/9801271].

\bibitem{SUSYFCNC}
See, for example,
  F.~Gabbiani, E.~Gabrielli, A.~Masiero and L.~Silvestrini,
  Nucl.\ Phys.\ B {\bf 477}, 321 (1996)
  [arXiv:hep-ph/9604387], and references therein.

\bibitem{Nomura-Yanagida}
  Y.~Nomura and T.~Yanagida,
  Phys.\ Lett.\ B {\bf 487}, 140 (2000)
  [arXiv:hep-ph/0005211].

\bibitem{gravitino-problem1}
  M.~Y.~Khlopov and A.~D.~Linde,
  Phys.\ Lett.\ B {\bf 138}, 265 (1984);
%
  J.~R.~Ellis, J.~E.~Kim and D.~V.~Nanopoulos,
  Phys.\ Lett.\ B {\bf 145}, 181 (1984).

\bibitem{gravitino-problem2}
For the latest analysis, see
  K.~Kohri, T.~Moroi and A.~Yotsuyanagi,
  arXiv:hep-ph/0507245,
  and references therein.

\bibitem{Nomura-Suzuki}
  Y.~Nomura and K.~Suzuki,
  Phys.\ Rev.\ D {\bf 68}, 075005 (2003)
  [arXiv:hep-ph/0110040].

\bibitem{WMAP}
 D.~N.~Spergel {\it et al.}  [WMAP Collaboration],
 Astrophys.\ J.\ Suppl.\  {\bf 148}, 175 (2003)
 [arXiv:astro-ph/0302209].

\bibitem{GNO2}
 H.S.~Goh, S.P.~Ng and N.~Okada, in preparation.

\bibitem{radioncosmo}
  C.~Cs\'{a}ki, M.~Graesser, L.~Randall and J.~Terning,
  Phys.\ Rev.\ D {\bf 62}, 045015 (2000)
  [arXiv:hep-ph/9911406].

\end{thebibliography}
\end{document}